\documentclass[journal,12pt,onecolumn,draftclsnofoot]{IEEEtran}
\usepackage{}
 \usepackage{cite}
\usepackage{psfrag}
\usepackage{latexsym, amsmath, subfigure,color, amsfonts, amssymb,graphicx}
\usepackage{algorithm}
\usepackage{algorithmic}
\usepackage{url}
\usepackage{epstopdf}

\newtheorem{theorem}{Theorem}[section]

\newtheorem{corollary}[theorem]{Corollary}

\newtheorem{remark}[theorem]{Remark}

\newtheorem{definition}[theorem]{Definition}

\begin{document}

\title{Preserving Data-Privacy with Added Noises: Optimal Estimation and Privacy Analysis}

\author{Jianping He$^{1,2}$, Lin Cai$^2$ and Xinping Guan$^1$
\thanks{$1$: The Dept. of Automation, Shanghai Jiao Tong University, and the Key Laboratory of System Control and Information Processing, Ministry of Education of China, Shanghai, China {\tt\small{jianpinghe.zju@gmail.com}},  {\tt\small{xpguan@sjtu.edu.cn}}}
\thanks{$2$: The Dept. of Electrical \& Computer Engineering at the University of Victoria, BC, Canada {\tt\small{jphe@uvic.ca}}, {\tt\small{cai@ece.uvic.ca}}}
\thanks{The preliminary result of this work was submitted to IEEE Conference on Decision and Control (CDC), 2017.}
 }

\maketitle
\begin{abstract}
Networked systems often relies on distributed algorithms to achieve a global computation goal with iterative local information exchanges between neighbor nodes. To preserve data privacy, a node may add a random noise to its original data for information exchange at each iteration. Nevertheless, a neighbor node can estimate other's original data based on the information it received. The estimation accuracy and data privacy can be measured in terms of $(\epsilon, \delta)$-data-privacy, defined as the probability of $\epsilon$-accurate estimation (the difference of an estimation and the original data is within $\epsilon$) is no larger than $\delta$ (the disclosure probability).  How to optimize the estimation and analyze data privacy is a critical and open issue. In this paper, a  theoretical framework is developed to investigate how to optimize the estimation of neighbor's original data using the local information received, named optimal distributed estimation. Then, we study the disclosure probability under the optimal estimation for data privacy analysis.  We further apply the developed framework to analyze the data privacy of the privacy-preserving average consensus algorithm and identify the optimal noises for the algorithm.

\end{abstract}


\IEEEpeerreviewmaketitle

\section{{Introduction}} \label{Intro}

Without relying on a central controller, distributed algorithms are robust and scalable, so they have been widely adopted in networked systems to achieve global computation goals (e.g., mean and variance of the distributed data) with iterative local information exchanges between neighbor nodes~\cite{KarTIT14, nedictac15, Andreassontac14}.
In many scenarios, e.g., social networks, the nodes' original data may include users' private or sensitive information, e.g., age, income, daily activities, and opinions. With the privacy concern, nodes in the network may not be willing to share their real data with others. To preserve data privacy,  a typical method is adding random noises to the data to be released in each iteration.  With the noise adding procedure, the goal of  privacy-preserving distributed algorithms is to ensure  data privacy while achieving the global computation goal~\cite{Gulisano16, he16tacsubmit, huang15icdcn}.

Consensus, an efficient distributed computing and control algorithm, has been heavily investigated and widely applied, 
e.g., in distributed estimation and optimization~\cite{pasqualetti2010distributed,mateos2009distributed}, 
distributed energy management and scheduling~\cite{zhao2015tsg, hetsp15},  and time synchronization in sensor networks~\cite{schenato2011average, he2011times, carli14tac}. Recently, the privacy-preserving average consensus problem has attracted attention, aiming to guarantee that the privacy of the initial states is preserved while an average consensus can still be achieved \cite{ny14tac, Manitara13, huang12, Nozari16, yilin15tac}. The main solution is to add variance decaying and zero-sum random noises during each iteration of the consensus process.

In the literature, differential privacy, a formal mathematical standard, has been defined and applied
for quantifying to what extent individual privacy in a statistical database is preserved~\cite{Dwork06}. It aims to provide means to maximize the accuracy of queries from statistical databases while maintaining indistinguishability of its transcripts. To guarantee the differential privacy, a commonly used noise is Laplacian noise
~\cite{TIT16, TIT16ad}.

Different from the database query problems, for many distributed computing algorithms such as consensus, the key  privacy concern is to ensure that other nodes cannot accurately estimate the original data, instead of the indistinguishability. No matter what type of noise distribution is used,
there is a chance that an estimated value of the original data is close to the real data, such a probability cannot be directly measured by differential privacy. To quantify the estimation accuracy and data privacy,
we first define $\epsilon$-accurate estimation, i.e., the difference of the estimated value and the original data is no larger than  $\epsilon$.
We then define $(\epsilon, \delta)$-data-privacy in~\cite{he16tacsubmit} as that the probability of $\epsilon$-accurate estimation is no larger than $\delta$. 
Using the  $(\epsilon, \delta)$-data-privacy definition,
in this paper, we develop a theoretical framework to investigate how to optimize the estimation of neighbor's original data using the local information received, named optimal distributed estimation. Then, we study the disclosure probability under the optimal estimation for data privacy analysis.  The main contributions of this work are summarized as follows.
\begin{enumerate}
\item To the best of our knowledge, this is the first work to mathematically formulate and solve the optimal distributed estimation problem and data privacy problem for the distributed algorithm with a general noise adding mechanism. The optimal distributed estimation is defined as the estimation that can achieve the highest disclosure probability, $\delta$, of $\epsilon$-accurate estimation, given the available information set.
\item  A theoretical framework is developed to analyze the optimal distributed estimation and data privacy by considering the distributed algorithm with a noise adding procedure, where the closed-form solutions of both the optimal distributed estimation and the privacy parameter are obtained. The obtained results show that how the iteration process and the noise adding sequence affect the estimation accuracy and data privacy, which reveals the relationship among noise distribution, estimation and data privacy.
\item   We apply the obtained theoretical framework to analyze the privacy of a general privacy-preserving average consensus algorithm (PACA), and quantify the $(\epsilon, \delta)$-data-privacy of PACA. We also identify the condition that the data privacy may be compromised. We  further obtain the optimal noise distribution for  PACA under which the disclosure probability of $\epsilon$-accurate estimation is minimized, i.e., the highest  data privacy is achieved.
\end{enumerate}

The rest of this paper is organized as follows. Section~\ref{sec:pre} provides preliminaries and formulates the problem.  The optimal distributed estimation and the privacy analysis under different available information set are discussed in Sections~\ref{sec:mainresult1} and  \ref{sec:mainresult2}, respectively. In Section~\ref{sec:applicaiton}, we apply the framework to analyze the data privacy of PACA.  Concluding remarks and further research issues are given in Section~\ref{sec:conclusions}.

\section{Preliminaries Problem Formulation}\label{sec:pre}

A networked system is abstracted as an undirected and connected graph, denoted by $G = (V, E)$, where $V$ is the set of nodes and $E$ is the set of edges. An edge $(i, j)\in E$ exists if and only if (iff) nodes $i$ can exchange information with node $j$.  Let $N_i=\{j| (i, j)\in E \}$ be the neighbor set of node $i$ ($i\notin N_i$). Let $n=|V|$ be the total number of nodes and  $n \geq 3$. Each node $i$ in the network has an initial scalar state $x_i(0)\in\mathcal{R}$, which can be any type of data, e.g., the sensed or measured data of the node. Let $x(0)=[x_1(0), ...., x_n(0)]^T \in \mathcal{R}^n$ be the initial state vector.

  \begin{table}[t] \tabcolsep 1pt \caption{Important Notations} \vspace*{4pt} \centering \tabcolsep 0.5mm
    \begin{tabular}{c||l}
    \hline Symbol  &                   Definition\\ \hline
    $G$              &  the network graph\\
    $x_i(0)$              &  node $i$'s initial state\\
  $x(0)$              &  the initial state vector of all nodes\\
  $f_i(\cdot)$             & the distributed iteration algorithm\\
  $\Theta_i$    & the domain of random variable $\theta_i$\\
  $f_{\theta_i}(\cdot)$  & the PDF of random variable $\theta_i$\\
  $\mathcal{I}_{i}^{in}(k)$ & the noise input of node $i$ until iteration $k$\\
  $\mathcal{I}_{i}^{out}(k)$ & the information output of node $i$ until iteration $k$\\
  $\hat{x}_i^*(k)$ & the optimal distributed estimation of $x_i(0)$ until iteration $k$ \\
    $\epsilon$ & the measure on estimation accuracy\\
  $\delta$ & the disclosure probability\\
  $\mathcal{I}_{\nu}^{out} $ & the possible output when the initial input is $\nu$\\
  $\mathcal{I}_{j}^{i}(k)$ & the information available to node $j$ to estimate \\ &$x_i(0)$ until iteration $k$\\
        \hline
    \end{tabular}
\vspace*{-4pt}
\label{table:definitions}
\end{table}

\subsection{Privacy-Preserving Distributed Algorithm}
The goal of a distributed algorithm is to obtain the statistics of all nodes' initial states (e.g., the average, maximum, or minimum value, variance, etc.) in a  distributed manner. Nodes in the network use the local information exchange to achieve the goal, and thus each node will communicate with its neighbor nodes periodically  for data exchange and state update. With the privacy concerns, each node is not willing to release its real initial state to its neighbor nodes. A widely used approach for the privacy preservation is adding random noises  at each iteration for local data exchange.

Define $x_i^+(k)$  the data being sent out by node $i$ in iteration $k$, given by
\begin{align}\label{nod}
x_i^+(k)=x_i(k)+\theta_i(k),
\end{align}
where $\theta_i(k)\in \Theta_i$ is a random variable. When node $i$ receives the information from its neighbor nodes, it updates its state using the following function,
\begin{align}\label{da}
x_i(k+1)=f_i(x_i^+(k), x_j^+(k): j\in N_i),
\end{align}
where the state-transition function, $f_i: \mathcal{R}\times \mathcal{R}\times ... \times\mathcal{R} \rightarrow \mathcal{R}$, depends  on $x_i^+(k)$ and $x_j^+(k)$ for $j \in N_i $ only.
The above equation defines a distributed iteration algorithm with privacy preserving since only the neighbor nodes' information is used for state update in each iteration and the data exchanged have been mixed with random noises to preserve privacy. Hence, (\ref{da}) is named as a privacy-preserving distributed algorithm. Since the initial state is most important for each node in the sense of privacy, in this paper, we focus on the estimation and privacy analysis of nodes' initial states.

\subsection{Important Notations and Definitions}
Define the noise input and state/information output sequences of node $i$ in the privacy-preserving distributed algorithm until iteration $k$ by
\begin{equation}
\mathcal{I}_{i}^{in}(k)=\{\theta_i(0), ..., \theta_i(k)\},
\end{equation}
and
\begin{equation}
\mathcal{I}_{i}^{out}(k)=\{x_i^+(0), ...,  x_i^+(k)\},
\end{equation}
respectively. Note that for any neighbor node $j$, it can  not only receive the information output $I_i^{out}(k)$ of node $i$, but also eavesdrop the information output of all their common neighbor nodes, which means that there may be more information available for node $j$ to estimate $x_i(0)$ at iteration $k\geq 1$. Hence, we define
\begin{align*}
I_j^i(k)=&\{x_i^+(0), x_\ell^+(0), ...., x_i^+(k), x_\ell^+(k) ~|~ \\&~\ell=j~\text{or}~\ell\in N_i\cap N_j\},
\end{align*}
as the available information set/outputs for node $j$ to estimate $x_i(0)$  of node $i$ at iteration $k$.
Clearly, we have $I_i^{out}(0)=I_j^i(0)$ and $I_i^{out}(k)\subseteq I_j^i(k)$.

 Let $f_{\theta_i(k)}(z)$ be the probability density function (PDF) of random variable $\theta_i(k)$.  
Let $\mathcal{X}_i\subseteq \mathcal{R}$ be the set of the possible values of $x_i(0)$. Clearly, if $\mathcal{X}_i=\mathcal{R}$, it means that $x_i(0)$ can be any value in $\mathcal{R}$. Given any function $f(y)$, we define the function $f(y, \epsilon)$ as
\begin{align}
f(y, \epsilon)=f(y+\epsilon)-f(y-\epsilon),
\end{align}
and let
\begin{align}
\Omega_{f}^0=\{y | f(y, \epsilon)=0\}
\end{align}
be the zero-point set of $f(y, \epsilon)=0$. Let $\{\circ \}_b$ be the boundary point set of a given set $\{\circ \}$, e.g., $(0,1]_b=\{0, 1\}$.

Note that each node can estimate its neighbor nodes' initial states based on all the information it knows, i.e., the available information set of the node. For example,  based on $I_{j}^{i}(0)=x_i^+(0)$, node $j$ can take the probability over the space of  noise $\theta_i(0)$ (where the space is denoted
by $\Theta_i(0)$) to estimate the values of the added noises, and then infer the initial state of node $i$ using the difference between
 $x_i^+(0)$ and the real initial state $x_i(0)$, i.e., $\hat{x}_i(0)=x_i^+(0)-\hat{\theta}_i(0)$.  Hence, we give two definitions for the estimation as follows.
\begin{definition}\label{aede}
Let $\hat{x}_i$ be an estimation of variable $x_i$. If
$|x_i-\hat{x}_i|\leq\epsilon$, where $\epsilon\geq 0$ is a small constant, then we say $\hat{x}_i$ is an $\epsilon$-accurate estimation.
\end{definition}

Note that $\mathcal{I}_{i}^{out}(k)$ is the information output sequence of node $i$, which is related to $x_i(0)$ directly, and this should be considered in the estimation. Since only the local information is available to the estimation, we define the optimal distributed estimation of $x_i(0)$ as follows.

\begin{definition} \label{opestde}
Let $\mathcal{I}_{\nu}^{out}(k)$ be the possible output given the condition that $x_i(0)=\nu$ at iteration $k$. Considering $\epsilon$-accurate estimation, under $\mathcal{I}_j^i(k)$,
\begin{align*}
\hat{x}_i^*(k)=\arg\max_{\hat{x}_i \in \mathcal{X}_i} \Pr\left\{\mathcal{I}_{\nu}^{out}(k)=\mathcal{I}_{i}^{out}(k)\mid  \forall |\nu-\hat{x}_i|\leq \epsilon  \right\},
\end{align*}
is named the optimal distributed estimation of $x_i(0)$ at iteration $k$. Then,  $\hat{x}_i^*=\lim_{k\rightarrow \infty} \hat{x}_i^*(k)$ is named the optimal distributed estimation of $x_i(0)$.
\end{definition}


In order to quantify
the degree of the privacy protection of the privacy-preserving distributed algorithm and construct a relationship between estimation accuracy
and the privacy, we introduce the following  $(\epsilon, \delta)$-data-privacy definition.
\begin{definition}
A distributed randomized algorithm is $(\epsilon, \delta)$-data-private, iff
 \begin{equation}\label{deprivacy}
\delta=\Pr\{|\hat{x}_i^*-x_i(0)|\leq \epsilon\},
\end{equation}
where $\delta$ is the disclosure probability that the initial state $x_i(0)$  can be successfully estimated by others using the optimal distributed estimation in a given interval $[x_i(0)-\epsilon, x_i(0)+\epsilon]$.
\end{definition}

In the above definition, $\hat{x}_i^*$ depends on the output sequences, $\mathcal{I}_{i}^{out}(k)$, which are the functions of  random noise inputs $\mathcal{I}_{i}^{in}(k)$ and its neighbors' output $\mathcal{I}_{j}^{out}(k), j\in N_i$. All the possible outputs of $\mathcal{I}_{i}^{out}(k)$ under a privacy-preserving distributed algorithm should be considered to calculate $\delta$, and thus $\hat{x}_i^*$  is a random variable  in (\ref{deprivacy}).  There are two important parameters in the privacy definition, $\epsilon$ and $\delta$, where  $\epsilon$ denotes the estimation accuracy and $\delta$ is the disclosure probability ($\delta\leq 1$) denoting the degree of the privacy protection. A smaller value of $\epsilon$ corresponds to a higher accuracy, and a smaller value of $\delta$ corresponds to a lower maximum disclosure probability.

\subsection{Problem of Interests}
We have the following basic assumptions,  i) if there is no information of any variable $y$  in estimation, then the domain of $y$ is viewed as $\mathcal{R}$, ii) unless specified, the global topology information is
unknown to each node, iii) the initial states of nodes in the network are independent of each other, i.e., each node cannot estimate the other nodes' state directly based on its own state or the estimation is of low accuracy.

In this paper, we aim to provide a theoretical framework of the optimal distributed estimation and data privacy analysis for the privacy-preserving distributed algorithm (\ref{da}). Specifically, we are interesting in the following three issues: i) how to obtain the optimal distributed estimation and its closed-form expression considering the distributed algorithm (\ref{da}); 
ii) using the $(\epsilon, \delta)$-data-private definition to analyze the privacy of the distributed algorithm (\ref{da}), i.e., obtaining the closed-form expression of the disclosure probability $\delta$ and its properties; and iii) using the obtained theoretical results to analyze the privacy of the existing privacy-preserving average consensus algorithm, and finding the optimal noise adding process to the algorithm, i.e.,
\begin{equation}
 \begin{split}\label{problem:p1}
\min_{\mathcal{I}_{i}^{in}(\infty)}  & ~~ \delta.
\\ s.t. ~~ & \lim_{k\rightarrow \infty} x_i(k)=\bar{x},
\end{split}
\end{equation}
where $\bar{x}={\sum_{i=1}^n x_i(0)\over n}$ is the statistic goal, aiming at minimizing the disclosure probability while obtaining the average value of all initial states.

To solve the above issues, in the following,  we first consider the case that only the one-step information output ($I_{i}^{out}(0)=I_{j}^{i}(0)$), which depends on the initial state ($x_i(0)$) and the one-step noise ($\theta_i(0)$), is available, and obtain the optimal distributed estimation and privacy properties. This case is suitable for the general one-step random mechanism (e.g., \cite{he16titsubmit, Kargupta03}), and the theoretical results provide the foundations of the following analysis. Then, we consider the optimal distributed estimation under the information set $I_{j}^{i}(1)$, which reveals that how the iteration process affects the estimation and helps to understand the optimal distributed estimation under the information set $I_{j}^{i}(k)$ ($k\geq 1$). Based on the observations, we extend the results to the general case that $I_{j}^{i}(k)$ ($\forall k\geq 0$) is available for the estimation. Lastly, we apply the obtained results to the general PACA algorithm for privacy analysis, and discuss the optimal noises for preserving the data privacy.

\section{Optimal Distributed Estimation and Privacy Analysis under $I_{j}^{i}(0)$}\label{sec:mainresult1}
In this section, the optimal distributed estimation of $x_i(0)$ using the information $I_{j}^{i}(0)$ only is investigated, and the disclosure probability $\delta$ under the optimal estimation is derived.

\subsection{Optimal Distributed Estimation under $I_{j}^{i}0)$}\label{subsec:e0}
Let $e_{\theta_i(0)}$ be the estimation of $\theta_i(0)$ under $I_{j}^{i}(0)$.
The optimal distributed estimation of $x_i(0)$  under $I_{j}^{i}(0)$ and its closed-form expression are given in the following theorem.
\begin{theorem}\label{theorem1}
Considering the distributed algorithm (\ref{da}),  under $I_{j}^{i}(0)$, the optimal distributed estimation of $x_i(0)$ satisfies
\begin{align}\label{opex0}
\hat{x}_i^*(0) &=x_i^+(0)- e_{\theta_i(0)}(x_i^+(0)),
\end{align}
where
\begin{align}\label{etheta0}
e_{\theta_i(0)}(x_i^+(0))=\arg\max_{y \in \{x_i^+(0)-\mathcal{X}_i\}} \int_{y-\epsilon}^{y+\epsilon} f_{\theta_i(0)}(z) \text{d} z;
\end{align}
Specifically, if $\mathcal{X}_i=\mathcal{R}$, then
\begin{align}\label{opex0add}
\hat{x}_i^*(0) &=x_i^+(0)- e_{\theta_i(0)},
\end{align}
where
\begin{align}\label{etheta0s}
e_{\theta_i(0)}=\arg\max_{y \in \mathcal{R}} \int_{y-\epsilon}^{y+\epsilon} f_{\theta_i(0)}(z) \text{d} z,
\end{align}
which is independent of $x_i^+(0)$.
\begin{proof}
Given $I_{i}^{out}(0)$ and an estimation $\hat{x}_i(0)$, we have
\begin{align}
&\Pr\left\{\mathcal{I}_{\nu}^{out}(0)=\mathcal{I}_{i}^{out}(0)\mid  \forall |\nu-\hat{x}_i(0)|\leq \epsilon  \right\}\nonumber\\
=&\Pr\left\{\nu+\theta_i(0)=x_i^+(0)\mid  \forall |\nu-\hat{x}_i(0)|\leq \epsilon  \right\}
\nonumber\\
=&\int_{x_i^+(0)-\hat{x}_i(0)-\epsilon}^{x_i^+(0)-\hat{x}_i(0)+\epsilon} f_{\theta_i(0)}(z) \text{d} z.
\end{align}
From Definition \ref{opestde}, it follows that
\begin{align}\label{opx0e}
\hat{x}_i^*( 0)&=\arg\max_{\hat{x}_i(0) \in \mathcal{X}_i} \Pr\left\{\mathcal{I}_{\nu}^{out}(0)=\mathcal{I}_{i}^{out}(0)\mid \forall |\nu-\hat{x}_i(0)|\leq \epsilon  \right\} \nonumber \\
&= \arg\max_{\hat{x}_i(0) \in \mathcal{X}_i} \int_{x_i^+(0)-\hat{x}_i(0)-\epsilon}^{x_i^+(0)-\hat{x}_i(0)+\epsilon} f_{\theta_i(0)}(z) \text{d} z \nonumber \\
&=x_i^+(0)- \arg\max_{y \in \{x_i^+(0)-\mathcal{X}_i\}} \int_{y-\epsilon}^{y+\epsilon} f_{\theta_i(0)}(z) \text{d} z \nonumber \\
&=x_i^+(0)-e_{\theta_i(0)}(x_i^+(0)),
\end{align}
which concludes that (\ref{opex0}) holds.

If $\mathcal{X}_i=\mathcal{R}$, for any real number output of $x_i^+(0)$, we have
\[\{x_i^+(0)-\mathcal{X}_i\}=\{x_i^+(0)-\mathcal{R}\}=\mathcal{R}.\]
In this case, we have
\begin{align}\label{opex0sr}
&\arg\max_{y \in \{x_i^+(0)-\mathcal{X}_i\}} \int_{y-\epsilon}^{y+\epsilon} f_{\theta_i(0)}(z) \text{d} z \nonumber \\ =& \arg\max_{y \in \mathcal{R}} \int_{y-\epsilon}^{y+\epsilon} f_{\theta_i(0)}(z) \text{d} z.
\end{align}
Substituting (\ref{opex0sr}) into  (\ref{opx0e}) gives
\begin{align*}
\hat{x}_i^*( 0)&= x_i^+(0)- \arg\max_{y \in \mathcal{R}}  \int_{y-\epsilon}^{y+\epsilon} f_{\theta_i(0)}(z) \text{d} z \nonumber \\
&=x_i^+(0)-e_{\theta_i(0)},
\end{align*}
i.e., (\ref{opex0add}) holds. Thus, we have completed the proof.
\end{proof}
\end{theorem}

\begin{figure}[t]
\begin{center}
\subfigure[$\mathcal{X}_i\subset \mathcal{R}$]{\label{ex1}
\includegraphics[width=0.44\textwidth]{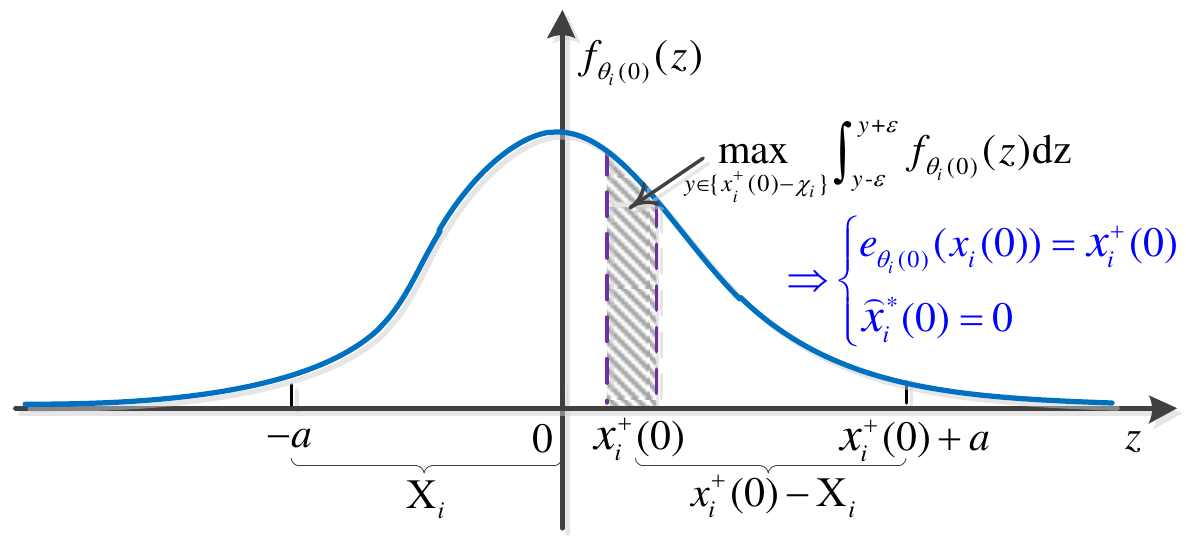}} 
\subfigure[$\mathcal{X}_i= \mathcal{R}$]{\label{ex2}
\includegraphics[width=0.45\textwidth]{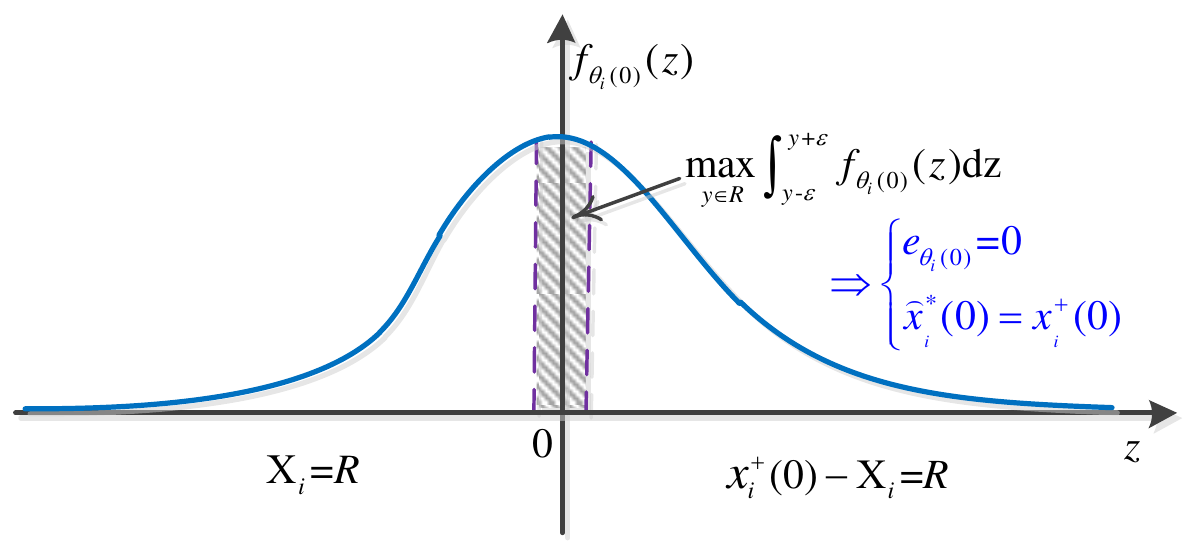}}
\caption{Two examples of the optimal distributed estimation under $I_{i}^{out}(0)$  considering $\mathcal{X}_i\subset \mathcal{R}$ and $\mathcal{X}_i= \mathcal{R}$, respectively.}\label{exam12}
\end{center}
\vspace*{-0pt}
\end{figure}

In (\ref{opex0}),  $e_{\theta_i(0)}(x_i^+(0))$ can be viewed as the estimation of the noise $\theta_i(0)$, i.e., $\hat{\theta}_i(0)=e_{\theta_i(0)}(x_i^+(0))$. Thus, (\ref{opex0}) can be written as
\begin{align*}
\hat{x}_i^*(0) &=x_i^+(0)- \hat{\theta}_i(0),
\end{align*}
which means that the estimation problem is equivalent to estimating the value of the added noise. From (\ref{etheta0}), it is noted that $e_{\theta_i(0)}(x_i^+(0))$  depends on $\epsilon$, $x_i^+(0)$, $f_{\theta_i(0)}$ and $\mathcal{X}_i$. We  use Fig. \ref{exam12}(a) as an example to illustrate how to obtain  $e_{\theta_i(0)}(x_i^+(0))$ and $\hat{x}_i^*(0)$ when $\mathcal{X}_i \subset \mathcal{R}$. Let the blue curve be the $f_{\theta_i(0)}(z)$ (it follows the Gaussian distribution in this example) and $\mathcal{X}_i=[-a, 0]$, and $x_i^+(0)$ is the fixed initial output. We then have
\[x_i^+(0)-\mathcal{X}_i=[x_i^+(0), x_i^+(0)+a].\]
Given an $\epsilon$ and $y\in [x_i^+(0), x_i^+(0)+a]$,  $\int_{y-\epsilon}^{y+\epsilon} f_{\theta_i(0)}(z) \text{d} z$ denotes the shaded area of $f_{\theta_i(0)}(z)$ in the interval $[y-\epsilon, y+\epsilon]$, which is named as the $\epsilon$-shaded area of $f_{\theta_i(0)}(z)$ at point $y$. Clearly, when $y=x_i^+(0)$, $f_{\theta_i(0)}(z)$ has the largest $\epsilon$-shaded area. It follows that $e_{\theta_i(0)}(x_i^+(0))=x_i^+(0)$, and thus $\hat{x}_i^*(0)=0$.
Meanwhile, we consider the case that  $\mathcal{X}_i=\mathcal{R}$ or $\mathcal{X}_i$ is not available to the other nodes, and use Fig. \ref{exam12}(b) as an example for illustration. In this case, we have $\mathcal{X}_i=x_i^+(0)-\mathcal{X}_i=\mathcal{R}$  for any output $x_i^+(0)$. From the above theorem,  we have \[e_{\theta_i(0)}(x_i^+(0))=e_{\theta_i(0)}=\arg\max_{y \in \mathcal{R}}  \int_{y-\epsilon}^{y+\epsilon} f_{\theta_i(0)}(z) \text{d} z=0.\]
Then, the optimal distributed estimation $\hat{x}_i^*(0)=x_i^+(0)-0=x_i^+(0)$ given any output $x_i^+(0)$.

Next, a general approach is introduced to calculate the value of $e_{\theta_i(0)}(x_i^+(0))$. Note that
\begin{align*}
{\partial ( \int_{y-\epsilon}^{y+\epsilon} f_{\theta_i(0)}(z) \text{d} z)
\over \partial y}& =F_{\theta_i(0)}(y+\epsilon)-F_{\theta_i(0)}(y+\epsilon)\nonumber \\
&= F_{\theta_i(0)} (y, \epsilon).
\end{align*}
It is well known that $F_{\theta_i(0)} (y, \epsilon)=0$ is a necessary condition that $y$ is an extreme point of $\int_{y-\epsilon}^{y+\epsilon} f_{\theta_i(0)}(z) \text{d} z$.
One then follows from (\ref{etheta0}) that $e_{\theta_i(0)}(x_i^+(0))$ is either one of the extreme points of $\int_{y-\epsilon}^{y+\epsilon} f_{\theta_i(0)}(z) \text{d} z$ (i.e., $e_{\theta_i(0)}(x_i^+(0))\in\Omega_{F_{\theta_i(0)}}^0$)  or one of the boundary points of $\{x_i^+(0)-\mathcal{X}_i\}$ (i.e., $e_{\theta_i(0)}(x_i^+(0)) \in\{x_i^+(0)-\mathcal{X}_i\}_b$. Let
 \[\mathcal{X}_{x_i^+(0)}=\{x_i^+(0)-\mathcal{X}_i\}\cap \Omega_{F_{\theta_i(0)}}^0\cup\{x_i^+(0)-\mathcal{X}_i\}_b,\]
 we then have
\begin{align}\label{et0cal}
e_{\theta_i(0)}(x_i^+(0))=\arg\max_{y \in\mathcal{X}_{x_i^+(0)}} \int_{y-\epsilon}^{y+\epsilon} f_{\theta_i(0)}(z) \text{d} z.
\end{align}

Applying the above general approach to the example of Fig. \ref{exam12}, one can easily obtain that
\[\mathcal{X}_{x_i^+(0)}=\{x_i^+(0), x_i^+(0)+a\}\]
and $\mathcal{X}_{x_i^+(0)}=\{0\}$ for the two cases, respectively. Based on (\ref{et0cal}), we obtain the same optimal estimations for the two cases.
\begin{remark}
From the above discussion, it is observed that $e_{\theta_i(0)}(x_i^+(0))$ is the point $y$  that  $f_{\theta_i(0)}(z)$ has the largest $\epsilon$-shaded area around point $y$, where $y\in \{x_i^+(0)-\mathcal{X}_i\}$. It should be pointed out that $e_{\theta_i(0)}(x_i^+(0))$ is in $\{x_i^+(0)-\mathcal{X}_i\}$ and depends on $\epsilon$, and thus it may not be the point that has the maximum value of $f_{\theta_i(0)}(z)$. However, if $\epsilon$ is sufficiently small and $f_{\theta_i(0)}(z)$ is continuous,  $f_{\theta_i(0)}(z)$ typically has the largest $\epsilon$-shaded area at point $y$ when $f_{\theta_i(0)}(y)$  has the maximum value for $y\in \{x_i^+(0)-\mathcal{X}_i\}$.  Meanwhile, the above examples also show that the unbiased estimation also may not be the optimal distributed estimation of $x_i(0)$.
\end{remark}

\subsection{Privacy Analysis  under $I_{j}^{i}(0)$}
In the above subsection, we have obtained the optimal distributed estimation when $I_{i}^{out}(0)$ is fixed.  Note that
\begin{align}\label{eeshi}
 &|\hat{x}_i^*(0)-x_i(0)|\leq \epsilon  \nonumber \\  \Leftrightarrow & |x_i^+(0)-x_i(0)-e_{\theta_i(0)}(x_i^+(0))|\leq \epsilon  \nonumber \\ \Leftrightarrow &|\theta_i(0)- e_{\theta_i(0)}(x_i^+(0))|\leq \epsilon
\end{align}
when $x_i^+(0)$ is fixed.  To analyze the privacy of  distributed algorithm   (\ref{da}) with the $(\epsilon, \delta)$-data-privacy definition, the main goal is to calculate the disclosure probability $\delta$, so that all the possible initial output  $x_i^+(0)$ and its corresponding optimal distributed estimation should be considered. Considering the outputs which can make an $\epsilon$-accurate estimations of $x_i(0)$ to be obtained,
we define all the corresponding noises by
\begin{align}
\mathcal{S}_i(0)=\{ \theta_i(0)  \mid   |e_{\theta_i(0)}(x_i^+(0)) -\theta_i(0)| \leq \epsilon\}.
\end{align}
For each $\theta_i(0)\in \mathcal{S}_i(0)$,  we have $x_i^+(0)=x_i(0)+\theta_i(0)$ and $ |e_{\theta_i(0)}(x_i^+(0)) -\theta_i(0)| \leq \epsilon$, i.e., an $\epsilon$-accurate estimation is obtained when $\theta_i(0)\in \mathcal{S}_i(0)$.

\begin{theorem}\label{theorem2}
Considering the distributed algorithm (\ref{da}), under $I_{j}^{i}(0)$,  the disclosure probability $\delta$ satisfies
\begin{align}\label{priacygeneralcase}
\delta=\oint_{\mathcal{S}_i(0)} f_{\theta_i(0)}(z) \text{d} z;
\end{align}
Specifically, if $\mathcal{X}_i=\mathcal{R}$, then
\begin{align}\label{priacygeneralcaser}
\delta=\max_{y \in \mathcal{R}} \int_{y-\epsilon}^{y+\epsilon} f_{\theta_i(0)}(z) \text{d} z.
\end{align}
\begin{proof}
From (\ref{eeshi}) and  the definition of $\delta$, we have
\begin{align}
\delta&= \Pr\{|\hat{x}_i^*(0)-x_i(0)|\leq \epsilon\} \nonumber
\\ &=\Pr\{|\theta_i(0)- e_{\theta_i(0)}(x_i^+(0))|\leq \epsilon\} \nonumber
\\ &=\Pr\{\theta_i(0)\in \mathcal{S}_i(0)\}  \nonumber
\\ &=\oint_{\mathcal{S}_{i}(0)} f_{\theta_i(0)}(z) \text{d} z.
\end{align}

From Theorem \ref{theorem1}, if $\mathcal{X}_i=\mathcal{R}$, then $e_{\theta_i(0)}(x_i^+(0))=e_{\theta_i(0)}$ which is independent of $x_i^+(0)$. In this case, we have
\begin{align}
\mathcal{S}_i(0)&=\{ \theta_i(0)  \mid  |e_{\theta_i(0)} -\theta_i(0)| \leq \epsilon\}\nonumber \\
&= [e_{\theta_i(0)} - \epsilon, e_{\theta_i(0)} + \epsilon],
\end{align}
i.e., only if $\theta_i(0)\in [e_{\theta_i(0)} - \epsilon, e_{\theta_i(0)} + \epsilon]$, we can obtain the $\epsilon$-accurate estimation of $x_i(0)$.
Then,
\begin{align}
\delta&= \oint_{\mathcal{S}_{i}(0)} f_{\theta_i(0)}(z) \text{d} z \nonumber
\\ &=\int_{e_{\theta_i(0)} - \epsilon}^{e_{\theta_i(0)} + \epsilon} f_{\theta_i(0)}(z) \text{d} z.
\end{align}
Since $e_{\theta_i(0)}$ satisfies (\ref{etheta0s}), $e_{\theta_i(0)}$ is the point $y$ that $f_{\theta_i(0)}(z)$ has the largest $\epsilon$-shaded area around it, and the domain of $y$ is $\mathcal{R}$. It follows that
\begin{align}
\delta&=\int_{e_{\theta_i(0)} - \epsilon}^{e_{\theta_i(0)} + \epsilon} f_{\theta_i(0)}(z) \text{d} z \nonumber
\\ &=\max_{y\in \mathcal{R}}\int_{y - \epsilon}^{y +\epsilon} f_{\theta_i(0)}(z) \text{d} z.
\end{align}
We thus have completed the proof.
\end{proof}
\end{theorem}

From the above theorem, we obtain that (\ref{priacygeneralcase}) provides the expression of the disclosure probability $\delta$ under $I_{i}^{out}(0)$.  Using (\ref{priacygeneralcase}), the main challenge to calculate $\delta$ is that how to obtain the set of $\mathcal{S}_i(0)$.  Although based on the definition of $\mathcal{S}_i(0)$, the elements of  $\mathcal{S}_i(0)$ can be obtained by comparing all possible values of $\theta_i(0)$ and their corresponding $e_{\theta_i(0)}(x_i^+(0))$ (how to obtain the value of $e_{\theta_i(0)}(x_i^+(0))$ is discussed in the previous subsection), this approach is infeasible  due to the infinite possible values of $\theta_i(0)$. Fortunately, we can apply the properties of $f_{\theta_i(0)}$ to fast obtain $\mathcal{S}_i(0)$ in many cases of practical importance. For the example given in Fig. \ref{exam12}(a),  since $f_{\theta_i(0)}$ is continuous and concave, it is straight-forward to obtain that
\begin{equation*}
e_{\theta_i(0)}(x_i^+(0))=\left\{ \begin{aligned}
       & x_i^+(0),  && x_i^+(0)\geq0;\\
        & 0, &&   x_i^+(0) \in [-a, 0]; \\
       & x_i^+(0)+a,  &&  x_i^+(0)\leq -a.
                          \end{aligned} \right.
                          \end{equation*}
Using the facts that $\hat{x}_i^*(0) =x_i^+(0)- e_{\theta_i(0)}(x_i^+(0))$ and $x_i^+(0)=x_i(0)+\theta_i(0)$, we then obtain
\begin{equation*}
\hat{x}_i^*(0)-x_i(0) =\left\{ \begin{aligned}
       & -x_i(0),  &&  x_i(0)+\theta_i(0)\geq0;\\
        & \theta_i(0), &&   x_i(0)+\theta_i(0)  \in [-a, 0]; \\
       & - a-x_i(0),  &&  x_i(0)+\theta_i(0) \leq -a.
                          \end{aligned} \right.
                          \end{equation*}
Based on the above equation, for any given $x_i(0)$ and $\epsilon$, we obtain all the $\theta_i(0)$  in $\mathcal{S}_i(0)$ by solving  $|\hat{x}_i^*(0)-x_i(0) |\leq \epsilon$, and thus $\mathcal{S}_i(0)$ is obtained.

\section{Optimal Distributed Estimation and Privacy  under $I_{j}^{i}(k)$}\label{sec:mainresult2}
In this section, we investigate the optimal distributed estimation and privacy under $I_{j}^{i}(1)$, and then extent the results to the general case that $I_{j}^{i}(k)$ is available to the estimation. Let $e_{\theta_i(0)|I_{j}^{i}(k)}$ be the estimation of $\theta_i(0)$ under $I_{j}^{i}(k)$.

\subsection{Optimal Distributed Estimation under $I_{j}^{i}(1)$}
Under $I_{j}^{i}(1)$, there are two outputs, $x_i^+(0)$ and $x_i^+(1)$, of node $i$, which can be used for initial state estimation or inference attack. Note that $x_i^+(1)=f_i(x_i^+(0), x_j^+(0): j\in N_i)$,  which means that $x_i^+(1)$ has involved the outputs of  node $i$'s neighbors. Hence, under $I_{j}^{i}(1)$, both the optimal distributed estimation and privacy analysis depend  on the output of both node $i$ and its neighbor nodes. 
Suppose that  $f_i$ in (\ref{da}) is available to the estimation in the remainder parts of this paper.

The following theorem provides the optimal distributed estimation of $x_i(0)$ under $I_j^i(1)$, which reveals the relationship between the information outputs (which are available to the node $j$ for estimation) and the optimal estimation.
 \begin{theorem} \label{theorem3}
Considering the distributed algorithm (\ref{da}),  under $I_{j}^{i}(1)$, the optimal distributed estimation of $x_i(0)$ satisfies
\begin{align}\label{opex0sd}
\hat{x}_i^*(1) &=x_i^+(0)- e_{\theta_i(0)|I_j^i(1)}(x_i^+(0)),
\end{align}
where
\begin{align}\label{ethetak}
e_{\theta_i(0)|I_j^i(1)}(x_i^+(0))& =\arg\max_{y \in \{x_i^+(0)-\mathcal{X}_i\}} \int_{y-\epsilon}^{y+\epsilon} \nonumber\\
&f_{\theta_i(1)}(\theta'_i(1))  f_{\theta_i(0)|\theta_i(1)=\theta'_i(1)}(z) \text{d} z,
\end{align}
in which $\theta'_i(1)=x_i^+(1)-f_i(x_i^+(0), x_j^+(0): j\in N_i)$; Then,
if $\mathcal{X}_i=\mathcal{R}$, we have
\begin{align}\label{ethetakad}
& e_{\theta_i(0)|I_j^i(1)}(x_i^+(0))= \arg\max_{y \in \mathcal{R}}\int_{y-\epsilon}^{y+\epsilon} \nonumber \\
&~~~~~~~~~~~ f_{\theta_i(1)}(\theta'_i(1))  f_{\theta_i(0)|\theta_i(1)=\theta'_i(1)}(z) \text{d} z.
\end{align}
\begin{proof}
Let  $\hat{x}_i(1)$ be an estimation of $x_i(0)$ under $I_{j}^{i}(1)$ at iteration $k=1$. Given $\mathcal{I}_{i}^{out}(1)$, and we have
\begin{align*}
&\Pr\{\mathcal{I}_{\nu}^{out}(1)=\mathcal{I}_{i}^{out}(1)\mid  \forall |\nu-\hat{x}_i(1)|\leq \epsilon, I_{j}^{i}(1) \}\nonumber\\
=&\Pr\{\mathcal{I}_{\nu}^{out}(1)=\{x_i^+(0), x_i^+(1)\}  \mid \forall~|\nu-\hat{x}_i(1)|\leq \epsilon, I_{j}^{i}(1)\} .
\end{align*}
Note that $x_i^+(0)$ depends on $x_i(0)$ and $\theta_i(0)$ only, while $x_i^+(1)$ depends on $x_i^+(0), x_j^+(0): j\in N_i$  and $\theta_i(1)$, where $\theta_i(0)$ and $\theta_i(1)$  are two random variables. It follows that
\begin{align}
&\Pr\{\mathcal{I}_{\nu}^{out}(1)=\{x_i^+(0), x_i^+(1)\}  \mid \forall~|\nu-\hat{x}_i(1)|\leq \epsilon,  I_{j}^{i}(1)\}\nonumber\\
=&\Pr\{\mathcal{I}_{\nu}^{out}(0, 0)=x_i^+(0),  \mathcal{I}_{i}^{out}(0, 1)=x_i^+(1) \mid \nonumber\\ &~~ \forall~|\nu-\hat{x}_i(1)|\leq \epsilon, {I_j^i(1)}\}
\nonumber\\
=& \int_{\hat{x}_i(1)-\epsilon}^{\hat{x}_i(1)+\epsilon} f_{\theta_i(0), \theta_i(1)}(x_i^+(0)-\nu, \theta'_i(1)) \text{d} \nu,
\end{align}
where
\[\theta'_i(1)=x_i^+(1)-f_i(x_i^+(0), x_j^+(0): j\in N_i).\]
Using the relationship between the joint distribution and the conditional distribution, one infers that
\begin{align}\label{condiae}
 &\int_{\hat{x}_i(1)-\epsilon}^{\hat{x}_i(1)+\epsilon} f_{\theta_i(0), \theta_i(1)}(x_i^+(0)-\nu, \theta'_i(1)) \text{d} \nu \nonumber\\
= &\int_{x_i^+(0)-\hat{x}_i(1)-\epsilon}^{x_i^+(0)-\hat{x}_i(1)+\epsilon} f_{\theta_i(1)}(\theta'_i(1))  f_{\theta_i(0)|{\theta_i(1)}}(z |\theta_i(1)=\theta'_i(1)) \text{d} z \nonumber\\
= & \int_{x_i^+(0)-\hat{x}_i(1)-\epsilon}^{x_i^+(0)-\hat{x}_i(1)+\epsilon}  f_{\theta_i(1)}(\theta'_i(1))  f_{\theta_i(0)| \theta_i(1)=\theta'_i(1)}(z) \text{d} z
\end{align}
where $f_{\theta_i(0)|{\theta_i(1)=\theta'_i(1)}}$ is the conditional PDF of $\theta_i(0)$ under the condition $\theta_i(1)=\theta'_i(1)$.
Then, one can obtain that
\begin{align}
&\max_{\hat{x}_i \in \mathcal{X}_i} \Pr\left\{\mathcal{I}_{\nu}^{out}(k)=\mathcal{I}_{i}^{out}(k) \mid  \forall |\nu-\hat{x}_i|\leq \epsilon,  I_{j}^{i}(1) \right\}\nonumber\\
= &\max_{\hat{x}_i \in \mathcal{X}_i} \int_{x_i^+(0)-\hat{x}_i(1)-\epsilon}^{x_i^+(0)-\hat{x}_i(1)+\epsilon}  f_{\theta_i(1)}(\theta'_i(1))  f_{\theta_i(0)| \theta_i(1)=\theta'_i(1)}(z) \text{d} z.
\end{align}
Hence,  we have
\begin{align*}
\hat{x}_i^*(1)=&\arg\max_{\hat{x}_i \in \mathcal{X}_i} \int_{x_i^+(0)-\hat{x}_i(1)-\epsilon}^{x_i^+(0)-\hat{x}_i(1)+\epsilon} \nonumber \\& ~~~f_{\theta_i(1)}(\theta'_i(1))  f_{\theta_i(0)| \theta_i(1)=\theta'_i(1)}(z) \text{d} z \nonumber \\
=&x_i^+(0)-e_{\theta_i(0)|I_j^i(k)}(x_i^+(0)).
\end{align*}

When $\mathcal{X}_i=\mathcal{R}$, we have $x_i^+(0)-\mathcal{X}_i=\mathcal{R}$ holds for any output $x_i^+(0)\in \mathcal{R}$.
It follows that (\ref{ethetak}) is equivalent to (\ref{ethetakad}) in this case. Thus,
the proof is completed.
\end{proof}
\end{theorem}

Note that the joint distribution of  any two random variables $X$ and $Y$ satisfies
\begin{align}\label{fact1}
f_{X, Y}(x, y)=f_{X|Y}(x|y)f_Y(y)=f_{Y|X}(y|x)f_X(x),
\end{align}
and we have
\begin{align}
 &\int_{\hat{x}_i(1)-\epsilon}^{\hat{x}_i(1)+\epsilon} f_{\theta_i(0), \theta_i(1)}(x_i^+(0)-\nu, \theta'_i(1)) \text{d} \nu \nonumber\\
= &\int_{x_i^+(0)-\hat{x}_i(1)-\epsilon}^{x_i^+(0)-\hat{x}_i(1)+\epsilon} f_{\theta_i(1)|{\theta_i(0)}}(\theta'_i(1)|\theta_i(0)=z)  f_{\theta_i(0)}(z) \text{d} z \nonumber\\
= & \int_{x_i^+(0)-\hat{x}_i(1)-\epsilon}^{x_i^+(0)-\hat{x}_i(1)+\epsilon}  f_{\theta_i(1)|{\theta_i(0)=z}}(\theta'_i(1))  f_{\theta_i(0)}(z) \text{d} z.
\end{align}
It thus follows that $e_{\theta_i(0)|I_j^i(1)}(x_i^+(0))$ also satisfies
\begin{align}\label{ethetakadequ}
& e_{\theta_i(0)|I_j^i(1)}(x_i^+(0)) \nonumber \\
=&\arg\max_{y \in \mathcal{R}}\int_{y-\epsilon}^{y+\epsilon}   f_{\theta_i(1)| \theta_i(0)=z}(\theta'_i(1))  f_{\theta_i(0)}(z) \text{d} z.
\end{align}
It should be noticed that $e_{\theta_i(0)|I_j^i(1)}(x_i^+(0))$ can be viewed as the optimal distributed estimation of $\theta_i(0)$ under $I_j^i(1)$, which depends on the distributions of $\theta_i(1)$ and $\theta_i(0)$,  the values of $x_i^+(0)$ and $\theta'_i(1)$, and $\mathcal{X}_i$.
Next, we  consider how these factors affect the value of $e_{\theta_i(0)|I_j^i(1)}(x_i^+(0))$.
\begin{corollary}\label{corollary1}
Considering the distributed algorithm (\ref{da}), if $\theta_i(0)$ and $\theta_i(1)$ are independent of each other,  under $I_{j}^{i}(1)$,  we have $e_{\theta_i(0)|I_j^i(1)}(x_i^+(0))=e_{\theta_i(0)}(x_i^+(0))$ and the optimal distributed estimation of $x_i(0)$ satisfies
\begin{align}\label{opex0sdc}
\hat{x}_i^*(1) &=\hat{x}_i^*(0)=x_i^+(0)- e_{\theta_i(0)}(x_i^+(0)).
\end{align}
\begin{proof}
For $e_{\theta_i(0)|I_j^i(1)}(x_i^+(0))$ in (\ref{opex0sd}), since $\theta_i(0)$ and $\theta_i(1)$ are independent of each other, we have that
\begin{align}
 f_{\theta_i(1)|\theta_i(0)}(\theta'_i(1)|\theta_i(0)=z)= f_{\theta_i(1)}(\theta'_i(1))
\end{align}
holds for $\forall z$.
Then, it follows from (\ref{fact1}) that
\begin{align}
& \int_{y-\epsilon}^{y+\epsilon}   f_{\theta_i(1)}(\theta'_i(1))  f_{\theta_i(0)| \theta_i(1)=\theta'_i(1)}(z) \text{d} z\nonumber \\
=&\int_{y-\epsilon}^{y+\epsilon}  f_{\theta_i(1)|\theta_i(0)}(\theta'_i(1)|\theta_i(0)=z)  f_{\theta_i(0)}(z) \text{d} z \nonumber \\
=&\int_{y-\epsilon}^{y+\epsilon}  f_{\theta_i(1)}(\theta'_i(1))  f_{\theta_i(0)}(z) \text{d} z \nonumber \\
=& f_{\theta_i(1)}(\theta'_i(1)) \int_{y-\epsilon}^{y+\epsilon}  f_{\theta_i(0)}(z) \text{d} z,
\end{align}
where $f_{\theta_i(1)}(\theta'_i(1)) $ is a constant when $I_{j}^{i}(1)$ is fixed. Together with  (\ref{ethetak}), one infers that
\begin{align}\label{e1e0e}
&e_{\theta_i(0)|I_j^i(1)}(x_i^+(0)) \nonumber \\
=& \arg\max_{y \in \{x_i^+(0)-\mathcal{X}_i\}}  \left(f_{\theta_i(1)}(\theta'_i(1)) \int_{y-\epsilon}^{y+\epsilon} f_{\theta_i(0)}(z) \text{d} z\right) \nonumber \\
=& \arg\max_{y \in \{x_i^+(0)-\mathcal{X}_i\}} \int_{y-\epsilon}^{y+\epsilon} f_{\theta_i(0)}(z) \text{d} z
\nonumber \\
=&
e_{\theta_i(0)}(x_i^+(0)),
\end{align}
where we use the fact that $f_{\theta_i(1)}(\theta'_i(1)) $ is a constant under $I_{j}^{i}(1)$.
From Theorem \ref{theorem3}, we have known that under $I_{j}^{i}(1)$, $\hat{x}_i^*(1)$ satisfies (\ref{opex0sd}). Substituting (\ref{e1e0e}) into  (\ref{opex0sd}), one obtains (\ref{opex0sdc}), which completes the proof.
\end{proof}
\end{corollary}

The above corollary shows that when the added noises are independent of each other,  the optimal distributed estimation $e_{\theta_i(0)|I_j^i(1)}(x_i^+(0))$  of $\theta_i(0)$ at iteration $k=1$ equals the optimal distributed estimation $e_{\theta_i(0)}(x_i^+(0))$  of $\theta_i(0)$ at iteration $k=0$, and thus we have $\hat{x}_i^*(1) =\hat{x}_i^*(0)$. Hence, one concludes that the later outputs cannot increase the estimation accuracy of $x_i(0)$ when the added noise sequence are independent of each other, and more details related to this conclusion will be provided in the next subsection.

\begin{corollary}\label{corollary22}
Considering the distributed algorithm (\ref{da}),  if $N_i\nsubseteq N_j$ for $\forall ~j\in N_i$ or the other nodes do not know all the information used for the updating by node $i$, under $I_{j}^{i}(1)$, the optimal distributed estimation of $x_i(0)$ satisfies
\begin{align}\label{gopex0sd}
\hat{x}_i^*(1) &=x_i^+(0)- e'_{\theta_i(0)|I_j^i(1)}(x_i^+(0)),
\end{align}
where
\begin{align}\label{gethetak}
e'_{\theta_i(0)|I_j^i(1)}(x_i^+(0))
=&\arg\max_{y \in \{x_i^+(0)-\mathcal{X}_i\}}\int_{y-\epsilon}^{y+\epsilon}\oint_{\Theta_{\theta'_i(1)|I_j^i(1)}} \nonumber \\&  f_{\theta_i(1)}(h)  f_{\theta_i(0)| {\theta_i(1)=h}}(z) \text{d} h \text{d} z,
\end{align}
and $\Theta_{\theta'_i(1)|I_j^i(1)}$ is the set of all possible values of $\theta'_i(1)$ under $I_j^i(1)$. Specifically, if $\Theta_{\theta'_i(1)|I_j^i(1)}\supseteq \Theta_i(1)$, we have $e'_{\theta_i(0)|I_j^i(1)}(x_i^+(0))=e_{\theta_i(0)}(x_i^+(0))$ and $\hat{x}_i^*(1) =\hat{x}_i^*(0)$.
\begin{proof}
For $\forall ~j\in N_i$,  since $N_i\nsubseteq N_j$,  there is at least one neighbor node of node $i$ satisfying $l\in N_i$ but $l\notin N_j$. It means that node $j$ cannot obtain all the neighbor nodes' information used for node $i$'s state updating. Thus, in the expression of $\theta'_i(1)$, there is at  least one unknown variable in $f_i(x_i^+(0), x_j^+(0): j\in N_i)$, which results that the exact value of $\theta'_i(1)$ cannot be obtained.
Hence, during the estimation, $\theta'_i(1)$  is no longer a deterministic value but is in a possible value set. Let  $\Theta_{\theta'_i(1)|I_j^i(1)}$ be set of the all possible values of $\theta'_i(1)$ under $I_j^i(1)$. During the estimation, we take all possible values of  $\theta'_i(1)$  into consideration, and then obtain \begin{align*}
&\Pr\{\mathcal{I}_{\nu}^{out}(1)=\{x_i^+(0), x_i^+(1)\} \mid \forall~|\nu-\hat{x}_i(1)|\leq \epsilon, I_i^j(1)\}\nonumber\\
= & \int_{x_i^+(0)-\hat{x}_i(1)-\epsilon}^{x_i^+(0)-\hat{x}_i(1)+\epsilon} \oint_{\Theta_{\theta'_i(1)|I_j^i(1)}}f_{\theta_i(1)} (h)  \text{d} h f_{\theta_i(0)|\theta_i(1)=h}(z) \text{d} z.
\end{align*}
Therefore, we have
\begin{align*}
\hat{x}_i^*(1)=&\arg\max_{\hat{x}_i \in \mathcal{X}_i}\Pr\{\mathcal{I}_{\nu}^{out}(1)=\{x_i^+(0), x_i^+(1)\} \mid\nonumber \\& ~~~\forall~|\nu-\hat{x}_i(1)|\leq \epsilon, I_i^j(1)\} \nonumber \\
=&x_i^+(0)-\arg\max_{y \in \{x_i^+(0)-\mathcal{X}_i\}}\int_{y-\epsilon}^{y+\epsilon}\oint_{\Theta_{\theta'_i(1)|I_j^i(1)}} \nonumber \\&  f_{\theta_i(1)}(h) \text{d} h  f_{\theta_i(0)|\theta_i(1)=h}(z)  \text{d} z \nonumber \\
=&x_i^+(0)-e'_{\theta_i(0)|I_j^i(k)}(x_i^+(0)).
\end{align*}

If $\Theta_{\theta'_i(1)|I_j^i(1)}\supseteq \Theta_i(1)$,  we have
\begin{align*}
&\int_{y-\epsilon}^{y+\epsilon}\oint_{\Theta_{\theta'_i(1)|I_j^i(1)}}   f_{\theta_i(1)}(h) \text{d} h  f_{\theta_i(0)|\theta_i(1)=h}(z)  \text{d} z\\ =& \int_{y-\epsilon}^{y+\epsilon}\oint_{\Theta_i(1)}  f_{\theta_i(1)|\theta_i(0)=z}(h)  \text{d} h  f_{\theta_i(0)}(z)  \text{d} z\\=& \int_{y-\epsilon}^{y+\epsilon}  f_{\theta_i(0)}(z)  \text{d} z,
\end{align*}
where we have used the fact that
\[\oint_{\Theta_i(1)}   f_{\theta_i(1)|\theta_i(0)=z}(h)  \text{d} h\equiv 1\]
holds for $\forall z\in \mathcal{R}$.
It thus has
\begin{align*}
e'_{\theta_i(0)|I_j^i(k)}(x_i^+(0))=&\arg\max_{y \in \{x_i^+(0)-\mathcal{X}_i\}}\int_{y-\epsilon}^{y+\epsilon}  f_{\theta_i(0)}(z)  \text{d} z \nonumber \\=& e_{\theta_i(0)}(x_i^+(0)),
\end{align*}
which means that
\[\hat{x}_i^*(1) =x_i^+(0)-e_{\theta_i(0)}(x_i^+(0))=\hat{x}_i^*(0).\]
We thus have completed the proof.
\end{proof}
\end{corollary}

In the above corollary,  if the assumption that for an unknown variable, it can be any value in $\mathcal{R}$ for estimation and $f_i(x_i^+(0), x_j^+(0): j\in N_i)$ with domain $\mathcal{R}$, then we have $\Theta_{\theta'_i(1)|I_j^i(1)}=\mathcal{R}$ since there is at least one unknown variable in $f_i(x_i^+(0), x_j^+(0): j\in N_i)$. Then, (\ref{gethetak}) can be simplified to
\begin{align*}
&e'_{\theta_i(0)|I_j^i(1)}(x_i^+(0))
 \nonumber \\=&\arg\max_{y \in \{x_i^+(0)-\mathcal{X}_i\}}\int_{y-\epsilon}^{y+\epsilon}\oint_{\mathcal{R}}  f_{\theta_i(1)}(h)  f_{\theta_i(0)| {\theta_i(1)=h}}(z) \text{d} h \text{d} z \nonumber \\= &\arg\max_{y \in \{x_i^+(0)-\mathcal{X}_i\}}\int_{y-\epsilon}^{y+\epsilon}\oint_{\mathcal{R}} f_{\theta_i(1)|{\theta_i(0)=z}}(h) \text{d} h  f_{\theta_i(0)}(z)  \text{d} z \nonumber
 \\= &\arg\max_{y \in \{x_i^+(0)-\mathcal{X}_i\}} f_{\theta_i(0)}(z)  \text{d} z,
\end{align*}
where we have used the facts that (\ref{fact1}) and
\[\int_{y-\epsilon}^{y+\epsilon}\oint_{\mathcal{R}} f_{\theta_i(1)|{\theta_i(0)=z}}(h) \text{d} h\equiv 1.\]

\begin{corollary}\label{corollary23}
Considering the distributed algorithm (\ref{da}), if $N_i\subseteq N_j$ and $N_i$ are known to node $j$,  under $I_j^i(1)$,  the optimal distributed estimation of $x_i(0)$ satisfies (\ref{opex0sd}) with
\begin{align*}
e_{\theta_i(0)|I_j^i(1)}(x_i^+(0))= \arg\max_{y \in \mathcal{R}}\int_{y-\epsilon}^{y+\epsilon}  f_{\theta_i(0)| \theta_i(1)=\theta'_i(1)}(z) \text{d} z.
\end{align*}
\begin{proof}
When $N_i\subseteq N_j$ and $N_i$ are known to node $j$, under $I_j^i(1)$,  node $j$ can obtain the exact value of $\theta'_i(1)$, since all the information of $x_i^+(1)-f_i(x_i^+(0), x_j^+(0): j\in N_i)$ are available to it. That is, $\theta'_i(1)$ is fixed when node $j$ makes the estimation, and thus
\begin{align}
& \arg\max_{y \in \{x_i^+(0)-\mathcal{X}_i\}}\int_{y-\epsilon}^{y+\epsilon} f_{\theta_i(1)}(\theta'_i(1))  f_{\theta_i(0)|\theta_i(1)=\theta'_i(1)}(z) \text{d} z
\nonumber \\=&\arg\max_{y \in \{x_i^+(0)-\mathcal{X}_i\}}\int_{y-\epsilon}^{y+\epsilon} f_{\theta_i(0)|\theta_i(1)=\theta'_i(1)}(z) \text{d} z.
\end{align}
Then, from Theorem \ref{theorem3}, we can obtain the results given in this corollary, which has completed the proof.
\end{proof}
\end{corollary}

 Corollaries \ref{corollary22} and \ref{corollary23} show the optimal distributed estimation of $x_i(0)$ until iteration $k$, considering node $j$ can and cannot have all information of the parameters in $f_i(x_i^+(0), x_j^+(0): j\in N_i)$ for the estimation, respectively. Therefore, they reveal that how the neighbor nodes' information outputs affect the optimal distributed estimation according to the iteration process of the privacy-preserving distributed algorithm (\ref{da}).

 \subsection{Optimal Distributed Estimation under $I_{j}^{i}(k)$}
 In this subsection, we consider the optimal distributed estimation of $x_i(0)$ under  $I_{j}^{i}(k)$. Let $k\rightarrow \infty$, and we obtain the optimal distributed estimation under $I_{j}^{i}(\infty)$.

We first give the following theorem,  which provides the expression of the optimal distributed estimation under  $I_{j}^{i}(k)$.
 \begin{theorem} \label{theorem3ik}
Considering the distributed algorithm (\ref{da}),  under $I_{j}^{i}(k)$, the optimal distributed estimation of $x_i(0)$ satisfies
\begin{align}\label{opex0sdik}
\hat{x}_i^*(k) &=x_i^+(0)- e_{\theta_i(0)|I_j^i(k)}(x_i^+(0)),
\end{align}
where
\begin{align}\label{ethetaik}
&e_{\theta_i(0)|I_j^i(k)}(x_i^+(0)) \nonumber \\
=&\arg\max_{y \in \{x_i^+(0)-\mathcal{X}_i\}}\int_{y-\epsilon}^{y+\epsilon}f_{\theta_i(1), ..., \theta_i(k)} (\theta'_i(1), ..., \theta'_i(k))\nonumber \\& f_{\theta_i(0)|\theta_i(k)=\theta'_i(k), ..., \theta_i(1)=\theta'_i(1)}(z) \text{d} z,
\end{align}
in which  $\theta'_i(k)=x_i^+(k)-f_i(x_i^+(k-1), x_j^+(k-1): j\in N_i)$.
\begin{proof}
From Theorem \ref{theorem3}, it is proved that (\ref{opex0sdik}) holds for $k=1$. Now, we prove that it holds for  $\forall k\geq 1$, where the basic idea is similar to the proof of Theorem \ref{theorem3}.

Let $\hat{x}_i(k)$ be an estimation of $x_i(0)$ under $I_{j}^{i}(k)$. We have the following equation holds,
\begin{align*}
&\Pr\left\{\mathcal{I}_{\nu}^{out}(k)=\mathcal{I}_{i}^{out}(k)\mid \forall |\nu-\hat{x}_i(k)|\leq \epsilon, I_{j}^{i}(k)\right\}\nonumber\\
=&\Pr\{\mathcal{I}_{\nu}^{out}(k)=\{x_i^+(0),  ..., x_i^+(k)\} \mid \forall~|\nu-\hat{x}_i(k)|\leq \epsilon, I_{j}^{i}(k)\}
\nonumber\\
=& \Pr\{\mathcal{I}_{\nu}^{out}(0, 0)=x_i^+(0),...,\mathcal{I}_{i}^{out}(k-1, k)=x_i^+(k)\mid \nonumber\\ & \forall~|\nu-\hat{x}_i(k)|\leq \epsilon, {I_j^i(k)}\}
\nonumber\\
=& \int_{\hat{x}_i(k)-\epsilon}^{\hat{x}_i(k)+\epsilon} f_{\theta_i(0), ..., \theta_i(k)}(x_i^+(0)-\nu, \theta'_i(1), ...., \theta'_i(k)) \text{d} \nu,
\end{align*}
where
\[\theta'_i(k)=x_i^+(k)-f_i(x_i^+(k-1), x_j^+(k-1): j\in N_i).\]
Using the properties of the joint distribution of multiple random variables, one infers that
\begin{align}
 & \int_{\hat{x}_i(k)-\epsilon}^{\hat{x}_i(k)+\epsilon} f_{\theta_i(0), ..., \theta_i(k)}(x_i^+(0)-\nu, \theta'_i(1), ...., \theta'_i(k)) \text{d} \nu  \nonumber\\
= &\int_{x_i^+(0)-\hat{x}_i(k)-\epsilon}^{x_i^+(0)-\hat{x}_i(k)+\epsilon}f_{\theta_i(1), ..., \theta_i(k)} (\theta'_i(1), ..., \theta'_i(k))\nonumber \\& f_{\theta_i(0)|\{\theta_i(1), ..., \theta_i(k)\}}(z | \theta'_i(1), ..., \theta'_i(k)) \text{d} z \nonumber\\
= &\int_{x_i^+(0)-\hat{x}_i(k)-\epsilon}^{x_i^+(0)-\hat{x}_i(k)+\epsilon}f_{\theta_i(1), ..., \theta_i(k)} (\theta'_i(1), ..., \theta'_i(k))\nonumber \\& f_{\theta_i(0)|\theta_i(k)=\theta'_i(k), ..., \theta_i(1)=\theta'_i(1)}(z) \text{d} z
\end{align}
where $f_{\theta_i(0)|\theta_i(k)=\theta'_i(k), ..., \theta_i(1)=\theta'_i(1)}(z)$ is the conditional PDF of $\theta_i(0)$  under the condition that $\{\theta_i(k)=\theta'_i(k), ..., \theta_i(1)=\theta'_i(1)\}$.
Then, one obtains that
\begin{align}
&\hat{x}_i^*(1)=\arg\max_{\hat{x}_i \in \mathcal{X}_i} \Pr\{\mathcal{I}_{\nu}^{out}(k)=\mathcal{I}_{i}^{out}(k)\mid \nonumber \\& ~~~ \forall |\nu-\hat{x}_i(k)|\leq \epsilon, I_{j}^{i}(k)\}\nonumber\\
&= \arg\max_{\hat{x}_i \in \mathcal{X}_i} \int_{x_i^+(0)-\hat{x}_i(k)-\epsilon}^{x_i^+(0)-\hat{x}_i(k)+\epsilon} f_{\theta_i(1), ..., \theta_i(k)} (\theta'_i(1), ..., \theta'_i(k))\nonumber \\&~~~ f_{\theta_i(0)|\theta_i(k)=\theta'_i(k), ..., \theta_i(1)=\theta'_i(1)}(z) \text{d} z \nonumber \\
&=x_i^+(0)-\arg\max_{y \in \{x_i^+(0)-\mathcal{X}_i\}}f_{\theta_i(1), ..., \theta_i(k)} (\theta'_i(1), ..., \theta'_i(k))\nonumber \\& ~~~ f_{\theta_i(0)|\theta_i(k)=\theta'_i(k), ..., \theta_i(1)=\theta'_i(1)}(z) \text{d} z \nonumber \\
&=x_i^+(0)-e_{\theta_i(0)|I_j^i(k)}(x_i^+(0)).
\end{align}
Thus,
the proof is completed.
\end{proof}
\end{theorem}

Then, we study the optimal distributed estimation of $x_i(0)$ under $I_{j}^{i}(k)$ and some other conditions, and provide three corollaries, respectively, as follows.
\begin{corollary}\label{corollary1k}
Considering the distributed algorithm (\ref{da}), if the added noises $\theta_i(0), ..., \theta_i(k)$ are independent of each other,  under $I_{j}^{i}(k)$,  $e_{\theta_i(0)|I_j^i(k)}(x_i^+(0))=e_{\theta_i(0)}(x_i^+(0))$ and the optimal distributed estimation of $x_i(0)$ satisfies
\begin{align}\label{opex0sdck}
\hat{x}_i^*(k) &=\hat{x}_i^*(0)=x_i^+(0)- e_{\theta_i(0)}(x_i^+(0)).
\end{align}
\begin{proof}
We only need to prove $e_{\theta_i(0)|I_j^i(k)}(x_i^+(0))=e_{\theta_i(0)}(x_i^+(0))$, then (\ref{opex0sdck}) can be inferred from Theorem \ref{theorem3ik} directly.   Since the added noises are independent of each other, we have
\[f_{\theta_i(0)|\theta_i(k)=\theta'_i(k), ..., \theta_i(1)=\theta'_i(1)}(z) =f_{\theta_i(0)}(z). \]
Then, (\ref{ethetaik}) can be simplified
\begin{align}\label{sethetaik}
&e_{\theta_i(0)|I_j^i(k)}(x_i^+(0))  \nonumber \\
=&\arg\max_{y \in \{x_i^+(0)-\mathcal{X}_i\}}f_{\theta_i(1), ..., \theta_i(k)} (\theta'_i(1), ..., \theta'_i(k)) \nonumber \\& \int_{y-\epsilon}^{y+\epsilon} f_{\theta_i(0)}(z) \text{d} z\nonumber \\
=&\arg\max_{y \in \{x_i^+(0)-\mathcal{X}_i\}}  \int_{y-\epsilon}^{y+\epsilon} f_{\theta_i(0)}(z) \text{d} z
\nonumber \\
=&e_{\theta_i(0)}(x_i^+(0)),
\end{align}
which completes the proof.
\end{proof}
\end{corollary}

\begin{corollary}\label{corollary2k}
Considering the distributed algorithm (\ref{da}),  if $N_i\nsubseteq N_j$ for $\forall ~j\in N_i$ or the other nodes do not know all the information used for the updating by node $i$, under $I_{j}^{i}(k)$, the optimal distributed estimation of $x_i(0)$ satisfies
\begin{align}\label{gopex0sdk}
\hat{x}_i^*(k) &=x_i^+(0)- e'_{\theta_i(0)|I_j^i(k)}(x_i^+(0)),
\end{align}
where
\begin{align}\label{gethetak}
&e'_{\theta_i(0)|I_j^i(k)}(x_i^+(0))\nonumber \\
=&\arg\max_{y \in \{x_i^+(0)-\mathcal{X}_i\}}\int_{y-\epsilon}^{y+\epsilon}\oint_{\Theta_{\theta'_i(1)|I_j^i(1)}}\cdot\cdot\cdot\oint_{\Theta_{\theta'_i(k)|I_j^i(k)}} \nonumber \\&f_{\theta_i(1), ..., \theta_i(k)} (z_k, ...,z_1) f_{\theta_i(0)|\theta_i(k)=z_k, ..., \theta_i(1)=z_1}(z_0) \nonumber \\& \text{d} z_k \cdot\cdot\cdot \text{d} z_1 \text{d} z_0,
\end{align}
$\Theta_{\theta'_i(k)|I_j^i(k)}$ is the set of all possible values of $\theta'_i(0)$ under $I_j^i(k)$. Specifically, if $\Theta_{\theta'_i(\ell)|I_j^i(\ell)}\supseteq \Theta_i$ holds for $\ell=1,..., k$,  $e'_{\theta_i(0)|I_j^i(k)}(x_i^+(0))=e_{\theta_i(0)}(x_i^+(0))$ and $\hat{x}_i^*(k) =\hat{x}_i^*(0)$.
\begin{proof}
Similar to the proof of Corollary \ref{corollary22}, if $N_i\nsubseteq N_j$ for $\forall ~j\in N_i$, there always exists unknown variables in the calculation of   $\theta'_i(1), ..., \theta'_i(k)$. Hence, under $I_{j}^{i}(k)$,  in (\ref{ethetaik}), $\theta'_i(1), ..., \theta'_i(k)$  cannot be fixed as constants during the estimation. Taking all the possible values of $\theta'_i(1), ..., \theta'_i(k)$ into consideration for the estimation,  (\ref{ethetaik}) is written as (\ref{gethetak}).

When $\Theta_{\theta'_i(\ell)|I_j^i(\ell)}\supseteq \Theta_\ell$ holds for $\ell=1,..., k$, we have the following equation
\begin{align}
 &\oint_{\Theta_{\theta'_i(1)|I_j^i(1)}}\cdot\cdot\cdot\oint_{\Theta_{\theta'_i(k)|I_j^i(k)}}f_{\theta_i(1), ..., \theta_i(k)|\theta_i(0)} \nonumber\\ &~~~(z_k, ...,z_1|\theta_i(0)=z_0) \text{d} z_k \cdot\cdot\cdot \text{d} z_1 \nonumber \\ =&\oint_{\Theta_1}\cdot\cdot\cdot\oint_{\Theta_k}f_{\theta_i(1), ..., \theta_i(k)|\theta_i(0)} \nonumber\\ &~~~(z_k, ...,z_1|\theta_i(0)=z_0) \text{d} z_k \cdot\cdot\cdot \text{d} z_1 \equiv  1,
\end{align}
holds for $\forall z_0$.
It follows that
\begin{align}
&\int_{y-\epsilon}^{y+\epsilon}\oint_{\Theta_{\theta'_i(1)|I_j^i(1)}}\cdot\cdot\cdot\oint_{\Theta_{\theta'_i(k)|I_j^i(k)}} f_{\theta_i(1), ..., \theta_i(k)} (z_k, ...,z_1)\nonumber \\&~~ f_{\theta_i(0)|\theta_i(k)=z_k, ..., \theta_i(1)=z_1}(z_0)  \text{d} z_k \cdot\cdot\cdot \text{d} z_1 \text{d} z_0
\nonumber \\=& \int_{y-\epsilon}^{y+\epsilon}\oint_{\Theta_{\theta'_i(1)|I_j^i(1)}}\cdot\cdot\cdot\oint_{\Theta_{\theta'_i(k)|I_j^i(k)}} f_{\theta_i(0)} (z_0)  \nonumber \\&~~ f_{\theta_i(1), ..., \theta_i(k)|\theta_i(0)} (z_k, ...,z_1| \theta_i(0)=z_0) \text{d} z_k \cdot\cdot\cdot \text{d} z_1 \text{d} z_0\nonumber \\
=& \int_{y-\epsilon}^{y+\epsilon} f_{\theta_i(0)} (z_0)  \text{d} z_0 .
\end{align}
Thus,  (\ref{gethetak}) is equivalent to
\begin{align}
e'_{\theta_i(0)|I_j^i(k)}(x_i^+(0))&=\arg\max_{y \in \{x_i^+(0)-\mathcal{X}_i\}} \int_{y-\epsilon}^{y+\epsilon} f_{\theta_i(0)} (z_0)  \text{d} z_0\nonumber \\&=e'_{\theta_i(0)}(x_i^+(0)),
\end{align}
which completes the proof.
\end{proof}
\end{corollary}

Note that if all the information used in (\ref{da}) is available to node $j$ for estimation, then values of $\theta'_i(1),..., \theta'_i(k)$ are fixed and known to node $j$. From Theorem \ref{theorem3ik}, we obtain the following corollary directly.

\begin{corollary}\label{corollary4.8}
Considering the distributed algorithm (\ref{da}), if $N_i\subseteq N_j$ and $N_i$ is known to node $j$, under $I_j^i(k)$, then the optimal distributed estimation of $x_i(0)$ satisfies (\ref{opex0sdik}) with
\begin{align*}
e_{\theta_i(0)|I_j^i(k)}(x_i^+(0))= & \arg\max_{y \in \{x_i^+(0)-\mathcal{X}_i\}}\int_{y-\epsilon}^{y+\epsilon}
\\ &f_{\theta_i(0) | \theta_i(1)=\theta'_i(1), ..., \theta_i(k)=\theta'_i(k)}(z) \text{d} z.
\end{align*}
\end{corollary}

The above three corollaries are correspondingly similar to Corollarys \ref{corollary1} to \ref{corollary23}, respectively.

\subsection{Disclosure Probability under $I_{j}^{i}(k)$}
The information set that can ensure an accurate estimation is defined by
\begin{align}
\mathcal{S}_i(k)=&\{I_j^i(k)  \mid   |e_{\theta_i(0)|I_j^i(k)}(x_i^+(0)) -\theta_i(0)| \leq \epsilon\}.
\end{align}
Then, define $\mathcal{S}_i^1(k)$ be the set of the first element in $\mathcal{S}_i(k)$, i.e., all the possible $x_i^+(0)$ included in $\mathcal{S}_i(k)$.
\begin{align}
\mathcal{S}_i^0(k)=&\{\theta_i(0) \mid  x_i^+(0) \in \mathcal{S}_i^1(k)\}.
\end{align}
Clearly, we have $\mathcal{S}_i^1(k)=x_i(0)+\mathcal{S}_i^0(k)$

The following theorem provides an upper bounded of the disclosure probability under  $I_j^i(k)$, which is denoted by $\delta(k)$.
\begin{theorem}\label{theorem2ki}
Considering the distributed algorithm (\ref{da}),  the disclosure probability $\delta$ at iteration $k$ satisfies
\begin{align}
\delta (k) \leq  \oint_{\mathcal{S}_i^0(k)} f_{\theta_i(0)}(z) \text{d} z.
\end{align}
\begin{proof}
Given an $I_j^i(k)$, the optimal distributed estimation $\hat{x}_i^*(k)$ satisfies  (\ref{opex0sdik}). Then,
\begin{align}
 &|\hat{x}_i^*(k)-x_i(0)|\leq \epsilon  \nonumber \\  \Leftrightarrow & |x_i^+(0)-x_i(0)-e_{\theta_i(0)|I_j^i(k)}(x_i^+(0))|\leq \epsilon  \nonumber \\ \Leftrightarrow &|\theta_i(0)- e_{\theta_i(0)|I_j^i(k)}(x_i^+(0))|\leq \epsilon.
\end{align}
From the definition of $\delta$, we have
\begin{align}
\delta(k) &= \Pr\{|\hat{x}_i^*(k)-x_i(0)|\leq \epsilon\}\nonumber
\\ &=  \Pr\{|\theta_i(0)- e_{\theta_i(0)|I_j^i(k)}(x_i^+(0))|\leq \epsilon\}  \nonumber
\\ & =  \oint_{\mathcal{S}_{i}(k)} f_{I_j^i(k)}(z) \text{d} z,
\end{align}
where $f_{I_j^i(k)}(z)$ is the PDF of $I_j^i(k)$ (since $I_j^i(k)$ is random under the distributed algorithm).
From the above function, it is hard to calculate the value of $\delta$, since $f_{I_j^i(k)}(z)$ is unknown and difficult to  obtain  due to the coupled input random variables. However, note that for each $I_j^i(k)\in \mathcal{S}_{i}(k)$, its element $x_i^+(0)$ should satisfy $x_i^+(0)-x_i(0)=\theta_i(0)$ and $\theta_i(0)\in \mathcal{S}_i^0(k)$. It means that only if $\theta_i(0)\in \mathcal{S}_i^0(k)$, $|\theta_i(0)- e_{\theta_i(0)|I_j^i(k)}(x_i^+(0))|\leq \epsilon$ can be true. Thus,
\begin{align}
\delta(k)&= \oint_{\mathcal{S}_{i}(k)} f_{I_j^i(k)}(z) \text{d} z \nonumber
\\ &\leq  \oint_{\mathcal{S}_{i}^0(k)} f_{\theta_i(0)}(z) \text{d} z,
\end{align}
which completes the proof.
\end{proof}
\end{theorem}

If there exist other conditions for estimation, e.g., independent noise inputs,  we obtain the closed-form expression of $\delta$, and thus we have the following theorem.

\begin{theorem}\label{theorems0}
Considering the distributed algorithm (\ref{da}), under $I_{j}^{i}(k)$, if one of the following conditions holds,
\begin{enumerate}
\item the added noise sequence $\theta_i(0), ..., \theta_i(k)$ are independent of each other;
\item $\Theta_{\theta'_i(\ell)|I_j^i(\ell)}\supseteq \Theta_i$ or $\Theta_{\theta'_i(\ell)|I_j^i(\ell)}=\mathcal{R}$ holds for $\ell=1,..., k$ and $\forall k\geq1$;
\end{enumerate}
then $\delta(k)=\delta$ holds for $\forall k\geq 0$ and $\delta$ satisfies (\ref{priacygeneralcase}). Furthermore, if $\mathcal{X}_i=\mathcal{R}$, $\delta$ satisfies (\ref{priacygeneralcaser}).
\end{theorem}

The above theorem can be obtained from Corollaries \ref{corollary1k} and \ref{corollary2k} and Theorem \ref{theorem2}, so we omit its proof.

\subsection{Calculation of the Optimal Estimation}
From the discussions in the above subsections,  the optimal distributed estimation of $x_i(0)$ is the most important factor for the privacy analysis. We design an algorithm to calculate the optimal distributed estimation of $x_i(0)$ under $I_{j}^{i}(k)$ for $\forall k\geq0$. From Theorem \ref{theorem3ik},  one infers that the key challenge to obtain $\hat{x}_i^*(k)$ is to calculate $e_{\theta_i(0)|I_j^i(k)}(x_i^+(0))$. Similar to the general approach given in Sec. \ref{subsec:e0}, we design Algorithm~\ref{cofe}  to calculate $e_{\theta_i(0)|I_j^i(k)}(x_i^+(0))$.

\begin{algorithm} [h]
\caption{: Calculation of $e_{\theta_i(0)|I_j^i(k)}(x_i^+(0))$} \label{cofe}
{\small{
\begin{algorithmic}[1]
\STATE \textbf{Input:} the information $I_j^i(k)$,  the PDFs $f_{\theta_i(0)}(z), ..., f_{\theta_i(k)}$.
\STATE \textbf{Calculation:}  Using the correlation among $\theta_i(0), ..., \theta_i(k)$ to obtain joint PDF $f_{\theta_i(1), ..., \theta_i(k)} (\theta'_i(1), ..., \theta'_i(k))$ and the conditional PDF$f_{\theta_i(0)|\theta_i(k)=\theta'_i(k), ..., \theta_i(1)=\theta'_i(1)}$.
\STATE  Computing the following derivative to obtain $f'_{\theta}(y, \epsilon)$,
\begin{align}
 \frac{\partial \int_{y-\epsilon}^{y+\epsilon} f'_{\theta}(z) \text{d} z}{\partial y}=F'_{\theta}(y, \epsilon)
\end{align}
where
\begin{align}
f'_{\theta}(z)=&f_{\theta_i(1), ..., \theta_i(k)} (\theta'_i(1), ..., \theta'_i(k))
\nonumber\\&f_{\theta_i(0)|\theta_i(k)=\theta'_i(k), ..., \theta_i(1)=\theta'_i(1)}(z)
\end{align}
\STATE  Solving the following equation to obtain the zero point set, $\Omega_{F_{\theta}}^0$,
\begin{align}
F'_{\theta}(y, \epsilon)=0.
\end{align}
\STATE Calculating the set of
 \[\mathcal{X}_{x_i^+(0)}=\{x_i^+(0)-\mathcal{X}_i\}\cap \Omega_{f'_{\theta_i}}^0\cup\{x_i^+(0)-\mathcal{X}_i\}_b.\]
\STATE Obtaining the estimation by
\begin{align}
e_{\theta_i(0)|I_j^i(k)}(x_i^+(0))=\arg\max_{y \in\mathcal{X}_{x_i^+(0)}} \int_{y-\epsilon}^{y+\epsilon}  f'_{\theta}(z) \text{d} z.
\end{align}
\STATE   \textbf{Output:} the estimation of $e_{\theta_i(0)|I_j^i(k)}(x_i^+(0))$.
\end{algorithmic} } }
\end{algorithm}

\section{Case Studies and Optimal Noises}\label{sec:applicaiton}
Privacy-preserving average consensus algorithm (PACA) is a typical privacy-preserving distributed algorithm, which aims to guarantee that the privacy of the initial state is preserved and at the same time the average consensus can still be achieved \cite{ny14tac, Nozari16, yilin15tac}. The basic idea of PACA is adding and subtracting variance decaying and zero-sum random noises to the traditional consensus process. Differential privacy of PACA has been studied in \cite{Nozari16}. In this section, we focus on data privacy analysis of PACA.  We adopt the developed theories in the above section to analyze the $(\epsilon, \delta)$-data-privacy of the PACA algorithm, and then find the optimal noises for the algorithm to achieve the highest data privacy.
\subsection{Privacy of PACA}
Referring to the existing algorithms, we describe the PACA algorithm as follows:
\begin{align}\label{generalconsensus}
&x_i(k+1)=f_i(x_i^+(k), x_j^+(k): j\in N_i)\nonumber\\&=w_{ii}(x_i(k)+\theta_i(k))+\sum_{j\in N_i} w_{ij} (x_j(k)+\theta_j(k)),
\end{align}
for $\forall i\in V$ and $k\geq 0$, where $w_{ii}$ and $w_{ij}$ are  weights, and its matrix form is given by
\begin{align}\label{matrix_generalconsensus}
& x(k+1)=W(x(k)+\theta(k)), k\geq 0,
\end{align}
where $W\geq 0\in \mathcal{R}^{n\times n}$ is a doubly stochastic matrix satisfying
$w_{ii}>0$ and $w_{ij}>0$ for $(i,j)\in E$;  and each $\theta_i(k)\in \theta(k)$ satisfies $\mathbf{Var} \{\theta_i(k)\}<\varrho \mathbf{Var} \{\theta_i(k-1)\}$ (where $0<\varrho<1$) and $\sum_{k=0}^\infty\theta_i(k)=0$. When $\theta(k)=0$ for $k\geq 0$, it is proved in
\cite{Olshevsky11} that an average consensus is achieved by (\ref{matrix_generalconsensus}), i.e.,
\begin{align}\label{gcconvergence}
& \lim_{k\rightarrow \infty}x(k)= {\sum_{\ell=1}^n x_\ell(0)\over n} \mathbf{1}=\bar{x}.
\end{align}
When $\theta(k)\neq0$ for $k\geq 0$,  it is proved in \cite{yilin15tac} that an average consensus is achieved by (\ref{matrix_generalconsensus})  in the mean-square sense.

The following two theorems analyze the data privacy of the PACA algorithm under the conditions that node $j$ can and cannot have all the information used in the iteration process for the estimation, respectively.

\begin{theorem}\label{theoremc1}
If $N_i\subseteq N_j$  and $N_i$ is known to node $j$ for $j\in N_i$,  using PACA, we have $\delta=1$ holds for $\forall \epsilon>0$, i.e.,  $x_i(0)$ is  perfectly inferred.
\begin{proof}
 When $N_i\subseteq N_j$  and $N_i$ is known to node $j$ for $j\in N_i$, then the values of $\theta'_i(1), ..., \theta'_i(\infty)$ are fixed and released to node $j$ under $I_j^i(\infty)$. From Corollary \ref{corollary4.8}, it follows that
 \begin{align}\label{etinfty}
e_{\theta_i(0)|I_j^i(\infty)}&(x_i^+(0))=\arg\max_{y \in \{x_i^+(0)-\mathcal{X}_i\}}\int_{y-\epsilon}^{y+\epsilon}\nonumber
\\ & f_{\theta_i(0) | \theta_i(1)=\theta'_i(1), ..., \theta_i(\infty)=\theta'_i(\infty)}(z) \text{d} z.
\end{align}
Meanwhile, from $\sum_{k=0}^\infty \theta_i(k)=0$, it follows that $\theta_i(0)=-\sum_{k=1}^\infty \theta_i(k)$. Hence, in the right side of (\ref{etinfty}),    the maximum value of the integral is achieved when $y=-\sum_{k=1}^\infty \theta_i(k)$, i.e.,
 \begin{align*}
& e_{\theta_i(0)|I_j^i(\infty)}(x_i^+(0))=-\sum_{k=1}^\infty \theta'_i(k)=\theta_i(0).
\end{align*}
Then, we have
\begin{align*}
\hat{x}_i^*(\infty) &=x_i^+(0)- e_{\theta_i(0)|I_j^i(\infty)}(x_i^+(0))\\&=x_i^+(0)-\theta_i(0)=x_i(0),
\end{align*}
i.e., $x_i(0)$ is  perfectly inferred, and thus $\delta=1$.
\end{proof}
\end{theorem}

In the above proof,  Corollary  \ref{corollary4.8} is adopted to prove the theorem. Actually, if $\theta'_i(1), ..., \theta'_i(\infty)$ are fixed and released,  the values of $\theta_i(1), ..., \theta_i(\infty)$ are released to node $j$. Then, using the condition  $\sum_{k=0}^\infty \theta_i(k)=0$, we can obtain $\theta_i(0)$, and thus $x_i(0)$ is obtained from using $x_i^+(0)-\theta_i(0)=x_i(0)$. It obtains the same results as Theorem \ref{theoremc1}, and thus verifies the results of Corollary  \ref{corollary4.8}.

\begin{theorem}\label{theorems1}
If $N_i\nsubseteq N_j$ for $\forall ~j\in N_i$ and $\Theta_i(k)=\mathcal{R}$ for $\forall i\in V$, then the PACA algorithm achieves
$(\epsilon, \delta)$-data-privacy, where $\delta$ satisfies (\ref{priacygeneralcase}), and then if $\mathcal{X}_i=\mathcal{R}$, $\delta$ satisfies (\ref{priacygeneralcaser}).
\begin{proof}
Since the conclusion in this theorem are the same as Theorem \ref{theorems0}, we prove it by showing that one of the conditions in Theorem \ref{theorems0} holds.
 Since $N_i\nsubseteq N_j$ for $\forall ~j\in N_i$, which means that any neighbor node $j$ cannot obtain all the information using in the right-hand side of (\ref{generalconsensus}) at each iteration $k$. Hence, there exists at least one $x_{j_0}^+(k-1)$ ($j_0\neq j$ and $j_0\in N_i$) which is not available to node $j$ for estimation. Note that under the PACA algorithm,
 \begin{align*}
 \theta'_i(k)=x_i^+(k)-(w_{ii}x_i^+(k-1)+\sum_{j\in N_i} w_{ij} x_j^+(k-1)).
 \end{align*}
Since $x_{j_0}^+(k-1)=x_{j_0}(k-1)+\theta_{j_0}(k-1)$ and $\theta_{j_0}(k-1)\in \Theta_{j_0}(k-1)= \mathcal{R}$, we have $ \theta'_i(k)\in \mathcal{R}$ during the estimation, i.e., $\Theta_{\theta'_i(k)|I_j^i(k)}= \mathcal{R}$. Therefore, the second condition in Theorem \ref{theorems0} holds, and we thus have completed the proof.
\end{proof}
\end{theorem}

With the above theorem, it is not difficult to prove that the algorithms proposed in both  \cite{yilin15tac} and \cite{he16tacsubmit} provide
$(\epsilon, \delta)$-data-privacy and $\delta$ satisfies (\ref{priacygeneralcaser}).

\subsection{Optimal Noises}
In this subsection, we consider the optimization problem (\ref{problem:p1}). It is known that the noise adding process given in PACA can ensure that the average consensus can be achieved by the algorithm, which means that the constraint in (\ref{problem:p1}) is satisfied. Meanwhile, from Theorem \ref{theorems1}, it follows that  $\delta$ satisfies (\ref{priacygeneralcase}), when node $j$ cannot know all the information using in the consensus process at each iteration.  Thus, under PACA,  problem (\ref{problem:p1}) is equivalent to an unconstrained minimization problem as follows,
\begin{equation}
 \begin{split}\label{problem:p2}
\min_{f_{\theta_i(0)}(y)} ~~ & \delta=\oint_{\mathcal{S}_i(0)} f_{\theta_i(0)}(y) \text{d} y .
\end{split}
\end{equation}
In problem (\ref{problem:p2}),  there is no constraint on $f_{\theta_i(0)}(y)$ and it can be a PDF of  any distribution of noises. Hence, we can find a $f_{\theta_i(0)}(y)$ with a large variance such that $\delta$ is smaller than any given small value since $\mathcal{S}_i(0)$ is a bounded set. For example, when $\mathcal{X}_i=\mathcal{R}$, we have $\mathcal{S}_i(0) =[e_{\theta_i(0)} - \epsilon, e_{\theta_i(0)} + \epsilon]$. Then, a uniform distribution with $ f_{\theta_i(0)}(y)\leq {1\over M}$ ($M$ is a constant) can ensure that
\begin{align}
\delta=\oint_{\mathcal{S}_i(0)} f_{\theta_i(0)}(y) \text{d} y \leq {2\epsilon \over M}.
\end{align}
which means that $\delta$ can be an arbitrarily small value as $M$ can be set arbitrarily large. Hence, one concludes that, by adding uniformly distributed noises,  PACA can provide $(\epsilon, \delta)$-data-privacy with any small $\delta$.

Then, we consider the case that the variance of $\theta_i(0)$ is a constant. Note that a smaller $\epsilon$ means a higher accuracy estimation. It means that when $\epsilon$ becomes smaller, the value of $\delta$ is more important for the privacy preservation. Hence, we define the optimal distribution in the sense of the data-privacy as follows.
 \begin{definition} \label{definition3}
 We say $f_{\theta_i(0)}^*(y)$ is the optimal distribution of $\theta_i(0)$ for a PACA. If, for any given distribution $f_{\theta_i(0)}^1 (y)$, there exists an $\epsilon_1$ such that $\delta(f_{\theta_i(0)}^* (y), \epsilon)<\delta(f_{\theta_i(0)}^1 (y), \epsilon)$ holds for $\forall \epsilon\in (0, \epsilon_1]$.
 \end{definition}

Based on Definition \ref{definition3}, we formulate the following minimization problem,
\begin{equation}
 \begin{split}\label{problem:p4}
\min_{f_{\theta_i(0)} (y)} & ~~ \delta ,
\\ s.t. 
 ~~ & \mathbf{Var} \{\theta_i(0)\}=\sigma^2.
\end{split}
\end{equation}
From our previous research on this optimization problem~\cite{he16autosubmit}, the optimal solution is that the noise $\theta_i(0)$ should follow  a uniform distribution given $\epsilon\leq \sigma$.



\section{Conclusions}\label{sec:conclusions}

In this paper,  we have investigated the optimal distributed estimation
and privacy problem for privacy-preserving distributed algorithm. We  introduced the definition of the optimal distributed estimation and the $(\epsilon, \delta)$-data-privacy definition, which reveals the relationship between the privacy and the estimation accuracy. A theoretical framework was provided for the optimal distributed estimation and the privacy analysis, where both the
closed-form expressions of the optimal distributed estimation and the  privacy
parameters were obtained. With the obtained framework, we proved that the existing PACA algorithm is $(\epsilon, \delta)$-data-private and the optimal noises, which guarantees the minimized disclosure probability, was obtained. The applications of the proposed framework will be considered in our future work.

\end{document}